\documentstyle{memsait}
\input epsf.sty
\begin{opening}
\title{TWO {\it BEPPO}SAX OBSERVATIONS OF BL LAC} 
\vspace{-0.5cm}
\author{G. Tagliaferri$^1$, G. Ghisellini$^1$, M. Ravasio$^1$,
A. Celotti$^2$, M. Chiaberge$^2$, L. Chiappetti$^3$,
L. Costamante$^1$, P. Giommi$^4$, L. Maraschi$^5$, E. Massaro$^6$, 
R. Nesci$^6$, C.M. Raiteri$^7$, F. Tavecchio$^5$, G. Tosti$^8$, 
A. Treves$^9$, M. Villata$^7$, A. Wolter$^5$}
\institute{
$^1$ Osservatorio Astronomico di Brera, Merate, Italy;
$^2$S.I.S.S.A., Trieste, Italy;
$^3$Istituto di Fisica Cosmica, CNR, Milan, Italy;
$^4${\it Beppo}SAX SDC, Rome, Italy;
$^5$Osservatorio Astronomico di Brera, Milan, Italy;
$^6$Istituto Astronomico, Universit\`a La Sapienza, Rome, Italy;
$^7$Osservatorio Astronomico Torino, Pino Torinese, Italy;
$^8$Osservatorio Astronomico, Universit\`a di Perugia, Perugia, Italy; 
$^8$Universit\`a dell'Insubria, Como, Italy.}
\date{} 
\end{opening}

\begin{document}

\oddpagefooter{}{}{} 
\evenpagefooter{}{}{} 

\vspace{-0.3cm}

\begin{abstract}
We present the results of two ToO {\it Beppo}SAX observations
of BL\,Lac. During the first observation we detected both the 
synchrotron and the Compton components. Fast time variability 
was present, but {\bf only} for the synchrotron component.
During the second observation the spectrum was flatter and
only the Compton component was present. Four different SED are 
presented and can be described by a blob moving along a jet and 
responsible for the SSC emission. The seed photons for the 
Compton scattering are the synchrotron photons themself plus 
possibly external photons coming from the broad line region.
\end{abstract}


\section{Introduction}
BL lac objects are highly variable sources characterised by non thermal
emission that dominates from the radio to the $\gamma$-rays. 
The overall Spectral Energy Distribution (SED) is dominated by two broad 
emission peaks: the lower frequency peak is believed to be produced by
synchrotron emission, while the higher frequency peak is probably due 
to inverse Compton process. 
The variability for these sources is more pronounced at higher energies
and the overall SED can change drammatically during strong flare events
(e.g. Mkn\,501, Pian et al. 1997; 1ES 2344+514, Giommi et al. 1997).

Here we report on the two {\it Beppo}SAX observations of BL Lac itself,
during a campaign aiming to study other BL Lac objects while in a flaring state. 
BL Lac X-ray spectrum  is quite hard 
with an energy spectral index of {$\alpha = 0.4-0.9$}
(Sambruna et al. 1999; Madejski et al. 1999).
It has already been observed by {\it Beppo}SAX, in November 1997,
with a 2-10 keV flux of $\sim 2 \times 10^{-11}
\ {\rm erg \ s^{-1} \ cm^{-2}}$ (Padovani et al. in preparation).
This is comparable to the level observed by {\it Rossi}XTE during
the 1997, July outburst. Our two {\it Beppo}SAX observations were 
triggered by two optical flare events (R magnitude brighter than 13 mag),
although when the source was actually observed by {\it Beppo}SAX the
optical flux was lower (R=13.4-13.6 mag).

\begin{table}
\hspace{1.11cm} 
\begin{tabular}{|l|l|l|l|l|c|}
\hline
   & & & & 
\vspace{-3.3mm} \\
Date  & $\alpha_1$ &  $\alpha_2$ & F$_{\rm 2-10 keV}$  & $\chi^2_r$/d.o.f. \\
& & &  ergs cm$^{-2}$ s$^{-1}$  & \\
\hline
  & & & & 
\vspace{-3mm} \\
5-7 June 1999 &  $1.6 \pm 0.25$  & $0.15 \pm 0.22$ & 
 $0.6 \ \times10^{-11}$ & 1.03/122 \\  
  & & & &
 \vspace{-3mm} \\
5-6 December 1999 & $0.6^{-0.05}_{+0.04}$ &  & $1.2 \ \times10^{-11}$ & 1.02/130  \\
%
\hline

\end{tabular}

\centerline{N$_{\rm H}$ fixed at the value $2.5\times 10^{21} \ {\rm cm}^{-2}$.
The two power laws cross at $\sim 5.5$ keV.}

\end{table}

\section{Observation and Results}

On May, 1999, BL Lac was reported to be again in an optically bright state,
thus we triggered our {\it Beppo}SAX ToO and BL\,Lac was observed the
5-7 of June, 1999. However, the source was about a factor of
two weaker than the first {\it Beppo}SAX observation.
The 0.1-100 keV X-ray spectrum cannot be fitted by a single power law.
A broken power law gives a better fit with a concave shape
(see {Table 1}). For the interstellar absorption we find a value
of N$_{\rm H}=2.5 \times 10^{21} \ {\rm cm}^{-2}$, consistent with
the values found in previous works (see Sambruna et al. 1999;
Madejski et al. 1999). Thus, we probably detected 
both the synchrotron and the inverse Compton components.
During this observation we also find very short time
scale variability, with the {\bf 0.1-10 keV flux increasing by
a factor of two in about 20 minutes}. This variation was not detected in
the simultaneous R band light curve. This variability
is seen only in the synchrotron part of the spectrum.
In fact, both the LECS and the MECS instruments detect this
variability only below the break, $\sim 4$ keV (see Fig. 1
in Tagliaferri et al. 2000a).
This result is similar to what we observed during another
{\it Beppo}SAX ToO observation of the BL Lac object ON\,231
(Tagliaferri et al. 2000b).

\vspace{-0.4cm}
\begin{figure}[h]

\epsfysize=9cm 
\hspace{1cm}
\epsfbox{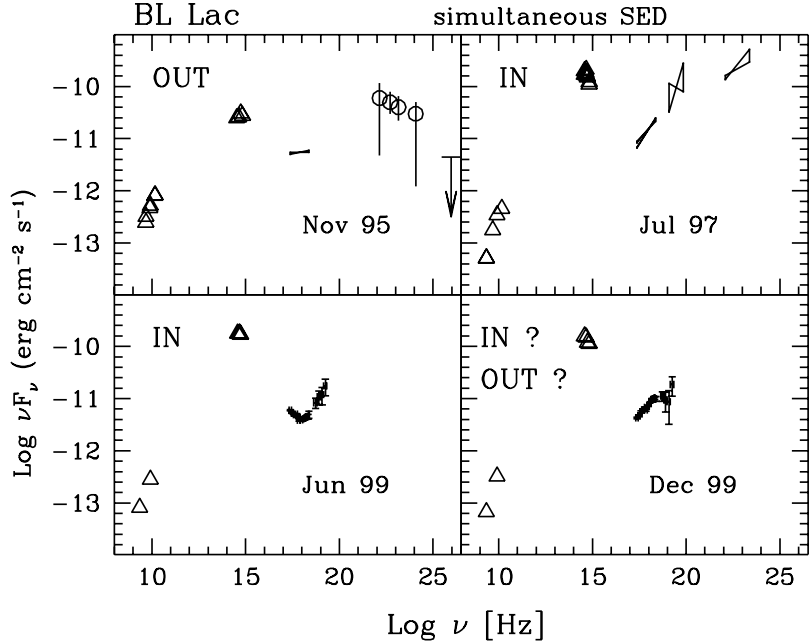} 
\vspace{-0.6cm}
\caption[h]{We plot four different SED of BL Lac. For the 1995 SED see
Sambruna et al. (1999), for the 1997 SED see  Madejski et al. (1999).
For the two {\it Beppo}SAX observations the simultaneous radio data
are from NRAO, while the optical data are from our observations.
The labels OUT or IN mean that the emitting blob is outside-inside
the BLR (see Discussion).}

\vspace{-0.2cm}
\end{figure}

After the {\it Beppo}SAX observation, the source remained active in
optical with the R magnitude ranging between $\sim 12.6-13.8$ mag.
At the end of November, it brightened again to R$\sim 12.4$, thus we 
triggered a second {\it Beppo}SAX ToO observation. However, again, 
when the source was observed in X-ray the R magnitude was down to
R$\sim 13.5$. This time the 0.1-100 keV X-ray spectrum is well fitted
by a single power law, with a flat energy index $\alpha = 0.6$.
The total flux in the 2-10 keV energy band is about a factor of two higher,
however the two spectra cross at about 1.5 keV, thus the flux
in the softer X-ray energy band is actually lower.
Clearly, the overall shape of the SED changed between the two observations
and during the second one we detected only the Compton component.
During this observation we did not detect fast variability.

\section{Discussion}

We plot in Fig. 1 four different SEDs corresponding to the multi-wavelength
campaigns carried out during November 1995, during
July 1997, when the source was in a very high state 
and during the two {\it Beppo}SAX observations. This figure clearly shows the
high degree of variability and complexity of the BL Lac's SED. Probably
different emission mechanisms are working at different times.
Here we propose the presence of a blob moving along a jet,
responsible for the SSC emission. If the moving blob is outside
the broad line region (BLR), there will not be an extra Compton component.
If instead the blob is inside the BLR, then there will also be external
seed photons coming from the BLR that can be Compton scattered.
These will be even more numerous if one or more clouds were
inside the jet itself. The various scenarios are sketched in Fig. 2.
In Fig. 1 the labels OUT/IN mean that the model that better represents
the data is the one with the blob INside or OUTside the BLR.

\begin{figure}[h]
\epsfysize=4.cm 
\epsfbox{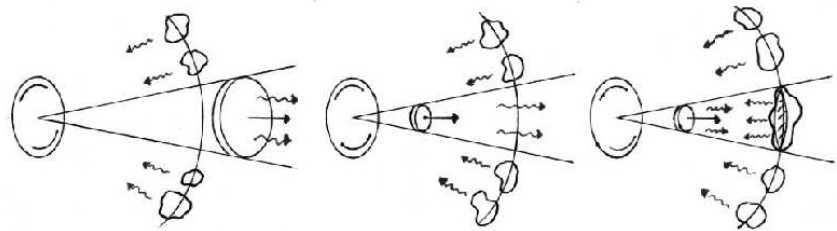} 
\caption[h]{We sketch here three different cases describing the 
position of the blob inside the jet with respect to the broad line region.
The different situations
can explain the SED observed at different epochs.}
\end{figure}


\end{document}